\documentclass[pre,twocolumn]{revtex4}
\usepackage{graphicx}
\usepackage{amsopn}
\usepackage{amsfonts}
\usepackage{amsmath}
\usepackage{latexsym}
\usepackage{amsmath,bm,amssymb}

\usepackage{color}

\begin{document}

\title{Probability of high intensities of the light wave \\
 propagating in turbulent atmosphere}

\author{I.V. Kolokolov, V.V. Lebedev}

\affiliation{Landau Institute for Theoretical Physics, Russian Academy of Sciences, \\
142432, Chernogolovka, Semenova 1A, Moscow region, Russia; \\
and National Research University  Higher School of Economics, \\
101000, Myasnitskaya 20, Moscow, Russia }

\begin{abstract}

We examine statistics of fluctuations of the laser beam intensity at its propagating in turbulent atmosphere. We are interested in relatively large propagating distances and the remote tail of the probability density function. The tail is determined by the stretched exponent, we find its index.

\end{abstract}


\maketitle

The subject of our work is the theoretical study of the physical properties of a laser beam propagating in a turbulent atmosphere or, more generally, an electromagnetic wave in a turbulent medium. The main effect that is being investigated in the framework of this problem is the diffraction of a laser beam by fluctuations of the refractive index, which are induced by pressure fluctuations of turbulent pulsations. The fluctuations of the refractive index are a random field whose properties are described statistically. Therefore, theoretical predictions of the behavior of the laser beam concern the average values (or the corresponding probability density functions), which are obtained by averaging over the statistics of fluctuations.

The problem of laser beam propagation in the atmosphere has a long history. The basic theoretical results concerning the propagation of a laser beam in a turbulent atmosphere were obtained back in the sixties and seventies of the last century, they are summarized in a number of monographs \cite{Tat71,Goodman,Tat75,StrohbehnBook1978,AndrewsPhillipsBook1998}. Recently, there has been a revival of interest in this topic \cite{Vorontsov2010,Vorontsov2011,Lachinova2016,Lushnikov2018,Fahey2021}, connected  mainly with numerical modeling of the propagation of a laser beam (electromagnetic wave) in a turbulent medium, which allows obtaining detailed information about the process.

The known theoretical results describe mainly the typical behavior of a laser beam in the atmosphere. At the same time, in the equation for the electromagnetic field, the random refractive index is a multiplicative noise, which leads to very non-trivial statistical properties of the electromagnetic field. It can be expected that the probability of rare events when one or another value (say, the intensity of the electromagnetic field) has an abnormally large value will be significantly higher than naive estimates based on the analysis of typical processes. As an example of such behavior, we can cite the statistics of quantum particles in a random potential (see, for example, \cite{quenched}), which is the multiplicative noise in the Schr\"odinger equation for a quantum particle.

We theoretically study the propagation of a monochromatic electromagnetic wave in an unlimited turbulent medium. All the characteristic scales of the problem (the size of the wave packet, the length of its propagation) are assumed to significantly exceed the wavelength, so that the description of the wave in terms of the complex envelope $\Psi$ is applicable. Due to the large value of the speed of light, we can assume that the state of the medium does not change during the propagation of the wave and use a stationary approximation to describe its envelope, that is, consider $\Psi$ as a function of the coordinate $z$ in the direction of wave propagation, and a two-dimensional radius vector ${\bm r}$ in the plane transverse to the direction of propagation. Of course, the state of the turbulent medium varies with time. In the accepted approximation, the envelope $\Psi(\bm r,z)$ adiabatically adjusts to the current state of the medium.

In this paper, we consider the intensity of the electromagnetic wave to be quite small, so that nonlinear effects (such as self-focusing) are insignificant. Then the equation for the envelope $\Psi(\bm r,z)$ in suitable units of measurement has the form of a two-dimensional Schr\"odinger equation
\begin{equation}
i\partial_z\Psi+ \nabla^2 \Psi +\xi \Psi=0,
\label{laser1}
\end{equation}
where $\nabla$ is the two-dimensional gradient in the plane transverse to the direction of propagation, and $\xi({\bm r},z)$ is the fluctuating component of the refractive index. By virtue of the equation (\ref{laser1}), the envelope $\Psi$ is a field that depends on the implementation of $\xi$. The latter varies over time, leading to variations in the envelope. Therefore, $\Psi$ can be considered as a random field whose statistical properties can be extracted by averaging over long times.

The distance traveled by the wave is assumed to be greater than the integral turbulence scale, whereas the transverse size of the wave packet is assumed to be smaller than the integral scale. In this case, the refractive index $\xi$ fluctuates rapidly along the direction of wave propagation. We are interested in the integral characteristics associated with $\xi$, therefore, by virtue of the central limit theorem, the random field $\xi$ can be assumed to have Gaussian probability distribution. It is determined by the pair correlation function of fluctuations in the refractive index:
\begin{equation}
\langle \xi(\bm r_1,z_1)\xi(\bm r_2,z_2) \rangle
=\delta(z_1-z_2)\left[\mathrm{const}-r_{12}^c\right],
\label{noise}
\end{equation}
where angle brackets mean averaging over implementations of the state of the medium, $r_{12}=|\bm r_1-\bm r_2|$ and $c$ is a number. The expression (\ref{noise}) is valid for distances $r_{12}$ lying in the inertial interval of turbulence. The constant in the ratio (\ref{noise}) is determined by turbulent fluctuations on the integral scale, and the power correction is determined by fluctuations on the scales $\sim r_{12}$. Fluctuations of the refractive index $\xi$ in a turbulent medium are determined mainly by pressure fluctuations. For the Kolmogorov spectrum, the exponent of $c$ in the expression (\ref{noise}) is $c=5/3$. Further, the exponent $c$ is treated as an arbitrary number lying in the interval $1 <c<2$.

We focus mainly on the statistical properties of the intensity $I$ of the laser beam (of the electromagnetic wave) at some observation point, which we choose as the origin: $I=|\Psi({\bm r}=0,z)|^2$. The distance $z$ traveled by the beam is assumed to be large, $z\gg1$ in our units. This means that the effects associated with random diffraction are strong. As the initial state, we choose a plane wave, $\Psi(\bm r,0)=1$. A generalization of our calculation scheme to other cases, say, to the initial Gaussian beam, will be published elsewhere. We will be interested in the moments of $\langle I^n\rangle$ for large values of $n$. These values determine the probability of events with an abnormally high intensity value.

The average value $\langle I^n \rangle$ can be expressed as $\langle I^n \rangle=F_{2n}(\bm 0, \dots , \bm 0,z)$, where $F_{2n}$ is the $2n$-point correlation function of the envelop:
 \begin{eqnarray}
 F_{2n}(\bm r_1, \dots , \bm r_{2n},z) = \qquad
 \nonumber \\
 \langle \Psi(\bm r_1,z) \dots
 \Psi(\bm r_n,z) \Psi^\star(\bm r_{n+1},z) \dots
 \Psi^\star(\bm r_{2n},z) \rangle.
 \label{cof1}
 \end{eqnarray}
This correlation function is represented as a convolution with the Green function ${\mathcal G}_{2n}$:
 \begin{eqnarray}
 F_{2n}=\int d^2x_1 \dots d^2x_{2n}
 {\mathcal G}_{2n}
 \Psi_{in}(\bm x_1) \dots \Psi_{in}^\star( \bm x_{2n}).
 \label{tord6}
 \end{eqnarray}
Here $\Psi_{in}(\bm r)=\Psi({\bm r},z=0)$ is the initial value of the envelop, and the Green function
\begin{equation}
{\mathcal G}_{2n} =
{\mathcal G}_{2n}(\bm r_1,\dots, \bm r_{2n},\bm x_1,\dots, \bm x_{2n},z),
\label{green}
\end{equation}
depends on $z$ and $4n$ radius-vectors.

The Green function can be represented as a path integral, which is derived from the Schr\"odinger equation (\ref{laser1}). After averaging over the fluctuations of the refractive index in accordance with the expression (\ref{noise}), we arrive at an integral over the variables $\bm y_j$, which are functions of the coordinate $\zeta$, $0<\zeta<z$:
\begin{eqnarray}
{\mathcal G}_{2n}=\int \prod_{j=1}^{2n}
D\bm y_j \exp\left\{\int_0^z d\zeta\,
\left(\frac{i}{4}K+W\right) \right\},
\label{puth1} \\
K= \left(\frac{d \bm y_1}{d\zeta}\right)^2 +\dots
+\left(\frac{d \bm y_n}{d\zeta}\right)^2 \qquad
\nonumber \\
-\left(\frac{d \bm y_{n+1}}{d\zeta}\right)^2
-\dots -\left(\frac{d \bm y_{2n}}{d\zeta}\right)^2, \qquad
\label{puth2} \\
 W={\sum_{i=1}^n}\sum_{j=i+1}^n y_{ij}^{c}
 +\sum_{i=n+1}^{2n} \sum_{j=i+1}^{2n} y_{ij}^{c}
 -\sum_{i=1}^n \sum_{j=n+1}^{2n}y_{ij}^{c}.
 \label{tord8}
\end{eqnarray}
One takes arguments $\bm x_j$ and $\bm r_j$ of the Green function as boundary conditions for the trajectories $\bm y_j(\zeta)$ at $\zeta=0$ and $\zeta=z$, see Eq. (\ref{green}). The ``potential'' $W$ (\ref{tord8}) is the function of the variables $y_{ij}=|\bm y_i-\bm y_j|$. The constant appearing in the expression (\ref{noise}) falls out of consideration, as it should be.

It is possible to find an explicit expression for the pair Green function ${\mathcal G}(\bm r_1,\bm r_2,\bm x_1,\bm x_2,z)$:
\begin{eqnarray}
 {\mathcal G}= \frac{\theta(z)}{16 \pi^2 z^2}
 \exp\left[\frac{i}{2z}(\bm r-\bm x)(\bm R-\bm X) \right.
 \nonumber \\
 \left.  -{z}\int_0^1 d\chi\, |\chi \bm x+(1-\chi)\bm r|^{c}  \right],
 \label{tord5}
 \end{eqnarray}
which determines the behavior of the pair correlation function. Here the designations $\bm R=(\bm r_1+\bm r_2)/2$, $\bm r=\bm r_1-\bm r_2$, $\bm X=(\bm x_1+\bm x_2)/2$, $\bm x=\bm x_1-\bm x_2$ are introduced.

For the initial state in the form of a plane wave, when $\Psi_{in}=1$, the pair correlation function has the simple form \cite{TatarskiiJETP1969, ZKT77}:
\begin{equation}
F(\bm r_1,\bm r_2,z)=\int d^2 x\, d^2 X {\mathcal G}=\exp(-z|\bm r_1-\bm r_2|^{c}).
 \label{turd9}
\end{equation}
Thus, the characteristic distance between the points $1$ and $2$ at a distance $z$ is $r_{ph}=z^{-1/c} \ll1$. This value makes sense of the phase corruption length due to fluctuations of the refractive index. The characteristic value of the variable $X$ is estimated as $z^{1/c+1}$.

The analysis of the $2n$-point Green function (\ref{green}) shows that for $z\gg1$ it has sharp maxima on configurations where there are $n$ pairs of close points $\bm x_j$ located at distances of the order of $r_{ph}$ separated by much larger distances \cite{ZKT77,KLL20}. In this case, the first point of the pair is taken from the set $\bm x_1, \dots, \bm x_n$, and the second is taken from the set $\bm x_{n+1}, \dots, \bm x_{2 n}$. The trajectories $\bm y_j(\zeta)$ in the path integral (\ref{puth1}) starting from close points $\bm x_j$ also remain close. In this case, the ``potential'' $W$ (\ref{tord8}) can be approximated by the sum of $y_{ij}^c$ for the pairs, and the remaining summands in the sum (\ref{tord8}) mutually reduce each other. Then the Green function is factorized. For example, for close couples $(\bm x_1,\bm x_{n+1})$, \dots $(\bm x_n,\bm x_{2n})$
\begin{eqnarray}
{\mathcal G}_{2n}=
{\mathcal G}(\bm r_1,\bm r_{n+1},\bm x_1,\bm x_{n+1},z)
\nonumber \\
\dots {\mathcal G}(\bm r_n,\bm r_{2n},\bm x_n,\bm x_{2n},z).
 \label{tord12}
\end{eqnarray}
Obviously, there is $n!$ of similar contributions, by the number of splits of $\bm x_j$ into pairs.

Thus, the space of integration by the initial coordinates in the expression for the moments of intensity
\begin{eqnarray}
\langle I^n \rangle
=\int d^2x_1 \dots d^2 x_{2n}
{\mathcal G}_{2n}(\bm 0, \dots, \bm 0, \bm x_1 \dots {\bm x}_{2n}).
\label{intensit2}
\end{eqnarray}
for the case of an initial plane wave is divided into $n!$ areas corresponding to different point splits into pairs. The contributions of all such regions are the same and in the main approximation for intensity moments we get $\langle I^n\rangle =n!$, which means the exponential probability density function $P(I)=\exp(-I)$. Note that $P(I)=\exp(-I)$ corresponds to the Gaussian statistics of the $\Psi$ field, natural for a complex field with a random phase.

It is demonstrated in the works \cite{ZKT77,KLL20,Char94} that corrections to the approximation (\ref{tord12}) related to the discarded terms in the expression (\ref{tord8}) for $W$ are proportional to the parameter $\alpha$:
\begin{equation}
\alpha = z^{-a}, \quad a=\frac{4}{c}-c,
\label{alpha}
\end{equation}
it is small at $z\gg1$. For the Kolmogorov spectrum of turbulence, the exponent $a$ is equal to $a=11/15$. Corrections to the value $\langle I^n \rangle=n!$ grow as $n$ increases and become essential when $\alpha n \sim 1$. In this paper, we determine the value of $\langle I^n\rangle$ at $\alpha n\gg 1$ and demonstrate that they significantly exceed $n!$. In other words, at $I\gg\alpha^{-1}$, the tail of the probability density function $P(I)$ arises, significantly exceeding $\exp(-I)$.

If $\alpha n \gg 1$ then the integration space in the integral (\ref{intensit2}) is still divided into $n!$ regions, corresponding to well separated pairs of the trajectories $\bm y_j(\zeta)$, the regions give identical contributions to $\langle I^n \rangle$. Let us take for definiteness the following subsetting into the pairs: ($\bm y_1,\bm y_{n+1}$), \dots ($\bm y_{n},\bm y_{2n}$). Introducing the variables $\bm Y_j=(\bm y_j+\bm y_{j+n})/2$ and $\bm\rho_j=\bm y_j-\bm y_{j+n}$, where $j$ runs through values from $1$ to $n$, we get a functional representation for the Green function with arguments in the selected area in the form:
\begin{eqnarray}
{\mathcal G}=
\int \prod_{j=1}^n D\bm Y_j\, D\bm \rho_j\,
\exp\left(-{\mathcal S}\right).
\label{path3}
\end{eqnarray}
In the path integral (\ref{path3}), some non-zero initial values are implied $\bm Y_j(0)=\bm X_j$ and zero final values (at $\zeta=z$) are implied for the trajectories $\bm Y_j$ and $\bm \rho_j$, the values are dictated by examining $\langle I^n \rangle$.

As it follows from the representation (\ref{puth1}), the action $\mathcal S$ in Eq. (\ref{path3}) is equal to
\begin{eqnarray}
{\mathcal S}=
-\frac{i}{2}\sum_j\int_0^z d\zeta\, \frac{d\bm Y_j}{d\zeta}\frac{d\bm \rho_j}{d\zeta}
+\sum_j\int_0^z d\zeta\, \rho_j^c
\nonumber \\
+\sum_{j>k}\int_0^z d\zeta\, U(\bm Y_j-\bm Y_k,\bm \rho_j, \bm \rho_k),
\label{action}
\end{eqnarray}
In this expression, the indices $j,k$ run through the values from $1$ to $n$. The quantity $U$ at $\bm \rho_j$, small in comparison with the characteristic values of $\bm Y_j-\bm Y_k$, is determined by the formula following from the expression (\ref{tord8}):
\begin{eqnarray}
U(\bm R,\bm \rho_1,\bm \rho_2)\approx \rho_{1,\alpha} \rho_{2,\beta}
V_{\alpha\beta}(\bm R),
\label{path5} \\
V_{\alpha\beta}(\bm R)
=c R^{c-2}\left[\delta_{\alpha\beta}
+(c-2) \frac{R_\alpha R_\beta}{R^2}\right].
\label{puth5}
\end{eqnarray}
Here the Greek indices $\alpha,\beta$ designate the components of the vectors in a plane transverse to the direction of wave propagation.

As we will see below, the values of $\bm Y_j$ in the integral (\ref{path3}) are parametrically large by $n$. This allows us to calculate the integral over $\bm Y_j$ in the expression (\ref{path3}) in the saddle-point approximation. The integral over $\bm\rho_j$ cannot be calculated in the approximation.

We first investigate the saddle-point value of $\bm Y_j$, which is determined by the extremum condition $\delta {\mathcal G}/\delta \bm Y_j=0$. Substituting here the expression (\ref{path3}), we find the equation
\begin{equation}
\frac{i}{2}\frac{d^2 \bar{\bm \rho}_j}{d\zeta^2}
+\sum_{k\neq j} \frac{\partial V_{\alpha\beta}}{\partial \bm Y_j}
(\bm Y_j-\bm Y_k)  \bar{\rho}_{j\alpha} \bar{\rho}_{k\beta}=0.
\label{aik7}
\end{equation}
Here $\bar {\bm \rho}_j$ is the average value of ${\bm \rho}_j$, it is non-zero because of non-zero initial values $\bm X_j$ of the trajectories $\bm Y_j$. When deriving the equation (\ref{aik7}), we replaced the average of the product of ${\rho}_{j\alpha}{\rho}_{k\beta}$ with the product of averages. The reason for this is that the fluctuations of $\bm\rho_j$ are determined mainly by the second term in the action (\ref{action}), which is diagonal by $j$. Therefore, the contribution of fluctuations to the mean ${\rho}_{j\alpha} {\rho}_{k\beta}$, where $j\neq k$, turns out to be negligible.

In field theory, the average value of the fluctuating field is found from the condition of the extremum of the so-called quantum effective action (see, for example, \cite{Weinberg}). In the problem under study, this condition is equivalent to the relation $\delta {\mathcal G}/\delta \bar{\bm \rho}_j=0$, which determines the average $\bar {\bm \rho}_j$. Utilizing the expression (\ref{path3}), we find the equation
\begin{eqnarray}
\frac{i}{2}\frac{d^2 Y_{j\alpha}}{d\zeta^2}
+\sum_{k\neq j} V_{\alpha\beta}(\bm Y_j-\bm Y_k)  \bar{\rho}_{k\beta}=0.
\label{aik8}
\end{eqnarray}
At deriving the equation we have neglected the contribution of the second term in the action (\ref{action}). The reason is that the fluctuations of $\bm\rho_j$ (which are determined mainly by this term) turn out to be much larger than $\bar{\bm\rho}_j$, which is why the dependence of the second term on $\bar{\bm\rho}_j$ is weak. Further, we will justify this neglect. Note that the extremum condition for the initial value of $\bar {\bm\rho}_j(0)$ leads to the condition $d \bm Y_j/d\zeta(0)=0$, due to the structure of the action (\ref{action}).

The equations (\ref{aik7},\ref{aik8}) show that the saddle-point value of the expression (\ref{path3}) is written as
\begin{eqnarray}
\ln{\mathcal G}_{sp}=\frac{i}{2}\sum_j\int_0^z d\zeta\, \frac{d\bm Y_j}{d\zeta}\frac{d \bar{\bm \rho}_j}{d\zeta}
\nonumber \\
-\sum_{j>k}\int_0^z d\zeta\, \bar\rho_{1,\alpha} \bar\rho_{2,\beta}
V_{\alpha\beta}(\bm Y_j-\bm Y_k)
\nonumber \\
=\sum_{j>k}\int_0^z d\zeta\, V_{\alpha\beta}(\bm Y_j-\bm Y_k)\bar{\rho}_{j,\alpha}\bar{\rho}_{k,\beta}.
\label{puth19}
\end{eqnarray}
The last equality in (\ref{puth19}) is obtained after integration in parts, taking into account the equation (\ref{aik8}) and the boundary conditions. The value (\ref{puth19}) is negative because $\bm Y_j$ are real quantities, whereas the averages $\bar{\bm\rho}_j$ are purely imaginary quantities, and the function $V_{\alpha\beta}(\bm Y_j-\bm Y_k)$ is positive for the values of the exponent $c$ considered here: $1<c<2$.

The system of equations (\ref{aik7},\ref{aik8}) enables one to evaluate the saddle-point quantity (\ref{puth19}). Assuming that all values of $Y_j$ together with their boundary values have the same order of $Y_j\sim X$, as well as the average $\bar{\rho}_j\sim i\bar{\rho}$, we arrive at the estimates
\begin{eqnarray}
\bar{\rho}\sim (nz^2)^{-1}X^{3-c}, \quad
\ln{\mathcal G}_{sp}\sim -\frac{1}{z^3} X^{4-c}.
\label{oik3}
\end{eqnarray}
Note the absence of dependence on $n$ in the estimation for $\ln{\mathcal G}_{sp}$.

Now we turn to accounting for the fluctuation contribution to ${\mathcal G}_{2n}$. Fluctuations of $\bm \rho_j$ can be estimated as $z^{-1/c}$, the value is much larger than the average $\bar\rho$ (\ref{oik3}). Therefore after shifting $\bm \rho_j$ by its mean value and shifting $\bm Y_j$ by its saddle-point value, the effects associated with the ``interaction potential'' (\ref{path5}) can be neglected (estimation of the accuracy see below). As a result, we arrive at a factorized approximation of the type (\ref{tord12}). Thus, the expression for the Green function is factorized ${\mathcal G}_{2n} ={\mathcal G}_{sp}{\mathcal G}_{fl}$, where ${\mathcal G}_{sp}$ is determined by the expression (\ref{puth19}) and the fluctuation factor is given by the product of the pair Green functions
\begin{eqnarray}
{\mathcal G}_{fl} =\prod_{j=1}^n
\frac{1}{16 \pi^2 z^2}
 \exp\left[ -\frac{z}{c+1} |\bm x_j-\bm x_{j+n}|^c\right].
\label{puth9}
\end{eqnarray}
The expression standing here for the pair Green functions is derived from the general formula (\ref{tord5}) after substituting $\bm r=\bm R=\bm 0$ and taking into account that the initial values of the shifted variables $\bm Y_j-\bm X_j$ are zero.

Returning to the expression (\ref{intensit2}) for the moments of intensity, we find
\begin{eqnarray}
\frac{\langle I^n \rangle}{n!}
=\int d^2x_1 \dots d^2 x_{2n}
{\mathcal G}_{sp} {\mathcal G}_{fl}.
\end{eqnarray}
The factor ${\mathcal G}_{sp}$ depends solely on the coordinates of the centers of the pairs $\bm X_j=(\bm x_j+\bm x_{j+n})/2$, therefore, after integration over differences $\bm x_j-\bm x_{j+n}$ we find
\begin{equation}
\frac{\langle I^n \rangle}{n!}\sim z^{-2(1+1/c)}
\int d^2 X_1 \dots d^2 X_n \, {\mathcal G}_{sp}.
\label{puth12}
\end{equation}
For a large $n$, the value of the integral (\ref{puth12}) is estimated as:
\begin{eqnarray}
\ln\frac{\langle I^n \rangle}{n!}\sim 2n \ln X
-2n\left(1+\frac{1}{c}\right) \ln z
-\frac{C}{z^3} X^{4-c},
\label{aik2}
\end{eqnarray}
where $C$ is a constant of the order of one. Here $X$ is the characteristic value of variables $\bm X_j$ introduced above.

Optimizing the expression (\ref{aik2}) by $X$, we find:
\begin{equation}
X^{4-c}\sim n z^3.
\label{aik3}
\end{equation}
The substitution of Eq. (\ref{aik3}) into Eq. (\ref{aik2}) gives the desired asymptotics of the high intensity moments:
\begin{eqnarray}
\ln \frac{\langle I^n \rangle}{n!} \sim
\frac{2}{4-c} n\ln n-\frac{2(4-c^2)}{(4-c)c}n \ln z.
\label{eik16}
\end{eqnarray}
The corresponding asymptotics of the probability density function $P(I)$ has the form
\begin{eqnarray}
\ln P\sim -\frac{1}{\alpha}(\alpha I)^{\beta}, \quad \beta=\frac{4-c}{6-c}.
\label{eik17}
\end{eqnarray}
For the Kolmogorov spectrum $\beta=7/13$. Since $\beta<1$, we conclude that there is a higher probability of large values of $I$ in comparison with the exponential probability density $\exp(-I)$. Formula (\ref{eik17}) is in accordance with the results of recent numerical simulations \cite{Lushnikov}. Note that the expression (\ref{eik17}) means a significant non-Gaussianity of the statistics of the wave field $\Psi$, formed by the chaotic scattering on the turbulent fluctuations.

The asymptotic law (\ref{eik16}) turns to $\langle I^n \rangle=n!$ at $\alpha n\sim 1$, or at $z\sim n^{c/(4-c^2)}$, as it should be. A similar statement is true for the probability density function, which is matched with the exponent $\exp(-I)$ at $I\sim 1/\alpha$. Eq. (\ref{aik3}) enables one to find the characteristic size of the region at the front of the initial wave defining $I^n$ at $n\gg 1/\alpha$:
\begin{equation}
X\sim (\alpha n)^{1/(4-c)}z^{1/c+1}.
\label{puth13}
\end{equation}
At $\alpha n \gg 1$ the quantity (\ref{puth13}) is much larger than the estimate $z^{1/c+1}$ for $X$ obtained in the Gaussian regime. Physically, this means that to create high intensity values, it is required to collect energy from a large area of the original wave. The expression (\ref{puth13}) implies an estimate for the average values $\bar{\rho}$ (\ref{oik3})
\begin{equation}
\bar\rho\sim \frac{1}{nz^2}X^{3-c}
\sim r_{ph}(n\alpha)^{-1/(4-c)}.
\label{puth14}
\end{equation}
These averages are small in comparison with the characteristic amplitude of fluctuations: $\bar{\rho}\ll r_{ph}$, which justifies the approximation (\ref{puth9}). Note, however, that the saddle-point action (\ref{puth19}) is determined precisely by these averages, and in action (\ref{puth19}) this smallness is compensated by a large number of terms.

The found asymptotic law (\ref{eik16}) is bounded from above by $n$. This limitation is determined by the mean square of the fluctuation component of the interaction of pairs between themselves that we have discarded. The assessment is fair for him
\begin{equation*}
z^2 n^2 r_{ph}^4 X^{2(c-2)}\sim n\left(n^{c/4}\alpha\right)^{4/(4-c)}.
\end{equation*}
The smallness of this value in comparison with the main value of the fluctuation action, estimated as $\sim n$, leads to the condition of applicability of our approach $n\ll\alpha^{-4/c}$. The inverse limit case requires a special analysis.

We have theoretically established that when a laser beam propagates in a turbulent medium at distances where diffraction on random fluctuations of the refractive index plays a significant role, the probability of abnormally large intensity fluctuations is significantly higher than estimates made on the basis of Gaussian statistics of the electromagnetic wave envelope $\Psi$. This is due to the multiplicativity of the refractive index in the equation for the envelop and it is a universal property of such stochastic systems. We have found the form of the probability distribution function of abnormally large intensity values. Our conclusions are made for the simplest case of the initial plane wave, although the analysis scheme itself is applicable to other initial forms of the laser beam. The results of their research will be published elsewhere. The case of extremely high intensity values requires a separate analysis. A separate analysis also requires to take into account the nonlinearity (for example, the effect of self-focusing), which may be significant for the large intensities of the electromagnetic wave we are considering.

The work is supported by the scientific program of the National Center for Physics and Mathematics, the project ``Physics of high energy densities'', stage 2023-2025. The authors thank P.M. Lushnikov for useful discussions.

\end{document}